\documentclass[aps, pre, twocolumn, floatfix, 10pt]{revtex4-2}
\usepackage{amsmath, amssymb, graphicx}
\usepackage{hyperref}
\hypersetup{colorlinks,allcolors=blue}

\begin{document}

\title{Influence of boundary geometry on active patterns}
\author{Jigyasa Watwani}
\author{Sakshi Pahujani}
\author{V Jemseena}
\author{Vishal Vasan}
\author{K Vijay Kumar}
\affiliation{International Centre for Theoretical Sciences, Tata Institute of Fundamental Research, \\ 
Survey 151, Shivakote Village, Hesaraghatta Hobli, 
Bengaluru North, India 560089.}

\date{\today}

\begin{abstract}
Mechanochemical patterns arising in the actomyosin cortex drive many cellular processes. Here we consider a hydrodynamic model for the actomyosin cortex of cells and study the sensitivity of the emergent patterns to both physical parameters and the geometry of the confining domain. We first establish a general framework for the Galerkin analysis of such patterns far from the linear stability regime on an arbitrary two dimensional domain. In the case of a circular disk, our analytical results predict transitions from isotropic to anisotropic patterns upon changing the strength of the active stress and the turnover rate. We confirm the existence of these genuine nonlinear bifurcations by an explicit numerical analysis of our model. Extending our numerical analysis to harmonic deformations of the circular disk, we show that the emergent patterns are also sensitive to the curvature of the domain. In particular, the actomyosin patterns resulting from our study closely resemble those seen in cells confined to micropatterned substrates. Our study demonstrates the role of geometry in controlling patterns within the context of a simple model for the actomyosin cortex.
\end{abstract}

\maketitle

\section{Introduction}
Pattern formation is a key driver of morphogenetic events at the cellular and developmental scales. The spatiotemporal variation of chemical concentration fields, flow fields, and filament orientation fields are implicated in many morphogenetic patterns. Many such patterns are mechanochemical in nature: chemical reactions and mechanical forces are fundamentally intertwined with each other. The origin of mechanical stresses can be traced to the forces generated by the energy consuming activity of molecular motors in the cellular cytoskeleton. Mechanochemical patterns in the actomyosin cytoskeleton are known to drive many cellular processes such as polarity establishment \cite{GoehringScience2011, GrossNatPhy2019}, cell division \cite{BrayWhite1988}, and cell motility \cite{BarrCell2007, MatsumuraTrendsCellBio2005, Pepper2022}. Even large scale shape deformations of tissues, such as wound healing and convergence-extension movements, are influenced by actomyosin patterns \cite{Salbreux2012, GrossAnnRevBiophysics2017, ClarkeCurrentBio2021}. 

The actomyosin cortex  -- a thin meshwork of actin filaments, myosin motors and associated cross-linking proteins located just beneath the plasma membrane -- is an excellent candidate for mechanochemical pattern formation \cite{Kumar2021}. Myosin motors use the energy released in ATP hydrolysis to tug antiparallel actin filaments, generating contractile active stresses in the process \cite{Kolomeisky2015motor}. Large scale gradients in these active stresses drive hydrodynamic flows \cite{MayerNature2010}. At high activity, advective fluxes can overcome the homogenizing effects of diffusion and lead to spatiotemporal patterns \cite{BoisPRL2011, KumarPRL2014}.

\begin{figure}[h]
\centering  \includegraphics[width=0.47\textwidth]{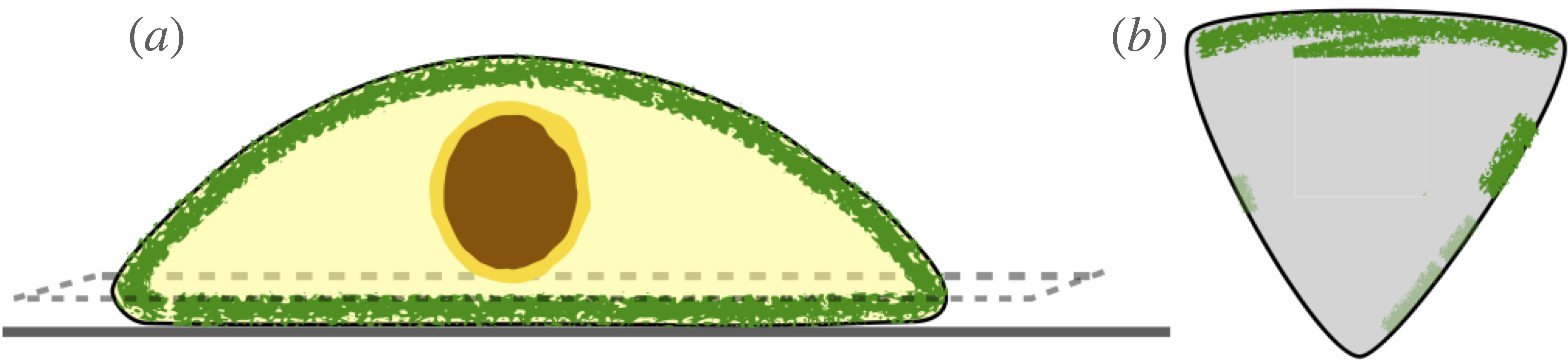 }
\caption{(a) The actoymosin cortex of cells is a thin film of active matter. (b) Concentration patterns seen in nearly flat cells laterally confined in micro-patterned substrates are sensitive to the geometrical shape of the confinement. This schematic figure is motivated by the results in \cite{TheryCellMotility2006,TheryJCellScience2010}.}
\label{fig:schematic}
\end{figure}

Theoretical studies of pattern formation typically consider a fixed geometry -- usually Cartesian domains with periodic or no-flux boundaries. On the other hand, morphogenetic patterns in cells and tissues display patterns on dynamical shape-shifting geometries. The dynamics of shape change is itself controlled by the mechanochemical stresses produced in the system. The resulting feedback between patterns and shape leads to very non-trivial problems. An intermediate case is to consider the regime where the timescales for shape change are longer than the timescales associated with the formation of patterns. In this quasi-static regime, one can study the effect of boundary shape on pattern selection. For reaction-diffusion dynamics, such as Turing systems, the role of boundary geometry in selecting patterns (even on dynamical domains with prescribed deformations) has been studied \cite{VenkatramanMainiPRE2011, PlazaJDynamicsDiffEqns2004, nampoothiri2017}. The evolution of parr-mark patterns from stripes to spots during the growth of Amago trout is an example of such a system \cite{VenkatramanMainiPRE2011}. However, the role of boundary geometry in controlling mechanochemical patterns has not been explored extensively. Experimentally, such mechanochemical patterns are seen in cells plated on micropatterned substrates with controlled boundary shape. For instance, cells adapt their actomyosin cortical patterns in response to the geometry of their confinement \cite{TheryJCellScience2010, TheryCellMotility2006, OakesBiophysicalJ2014, WangJMaterChemB2016} (see Fig~\ref{fig:schematic}). 

In this study, we consider a minimal mechanochemical model for the actomyosin cortex modeled as a thin film of an active material. The transport equation for the concentration field of an active stress regulator is coupled to the fluid flow resulting from active hydrodynamics. Considering finite-sized arbitrarily shaped two-dimensional domains with no-flux boundary conditions, we set up a general formalism to write an effective Galerkin truncated model for the dominant modes of the patterns. On a circular domain, our analytical predictions for the existence of anisotropically localized concentration patterns at the boundary, and their transition to  isotropic patterns with change of activity and turnover rate, is in very good agreement with explicit numerical simulations of the model equations. For geometries that are perturbations of the circular disk, we find that the patterns can spontaneously localize at regions of high or low boundary curvature depending on parameters. Our results on triangular shaped domains show remarkable similarity with experimentally observed patterns. The transitions between various patterns predicted by our model with change in geometry can be tested experimentally.

This manuscript is organized as follows: In Section \ref{sec:model}, we formulate the model and associated boundary conditions on arbitrary two-dimensional domains. In Section \ref{sec:framework}, we develop a general framework to obtain effective Galerkin truncated models of reduced dimensionality. This method works for any two-dimensional geometry in which the spectrum of the scalar Helmholtz equation can be obtained. In Section \ref{sec:results}, we first consider a circular domain and compare the analytical predictions of the Galerkin model with explicit numerical simulations of the full model. Next, we study patterns on two-dimensional shapes that are harmonic deformations of the circular disk. We then conclude with a discussion in Section \ref{sec:discussion}.

\section{Model}
\label{sec:model}
We consider the reaction-transport dynamics of a chemical regulator with concentration field $c$ confined to move in a thin film such as the actomyosin cortex of cells. The chemical regulator is transported by both diffusive and advective fluxes. In addition, a continuous exchange with the cytoplasm leads to a turnover of the material in the two-dimensional thin-film. The resulting equation for the dynamics of $c$ is
\begin{align}
\label{eqn_concentration}
\partial_t c = \mathcal{D} \, \nabla^2 c - \nabla \cdot (\boldsymbol{v} c) - \kappa \, (c - c_0),
\end{align}
where $\mathcal{D}$, $\kappa$ and $c_0$ are respectively the diffusion constant, turnover-rate, and base-level of $c$. Neglecting inertial effects, force-balance in the actomyosin cortex leads to an equation for the hydrodynamic velocity $\boldsymbol{v}$:
\begin{align}
\label{eqn_force_balance}
\nabla \cdot \boldsymbol{\mathsf{\Sigma}} = \gamma \, \boldsymbol{v},
\end{align}
where $\gamma$ is a friction coefficient and $\boldsymbol{\mathsf{\Sigma}}$ is the total hydrodynamic stress in the cortex. Our interest is in the patterns of $c$ observed on time-scales much longer than the Maxwell relaxation time ($\sim 5$s) of the cortex \cite{SAHA20161421}. As such, the total stress $\boldsymbol{\mathsf{\Sigma}}$ consists of a passive component arising from velocity gradients and an active component regulated by the concentration of the chemical regulator \cite{BoisPRL2011}. Furthermore, we neglect any orientational ordering of actin filaments and assume that the active stress is isotropic in the plane of the cortex. Therefore
\begin{align}
\label{eqn_total_stress}
\boldsymbol{\mathsf{\Sigma}} &= 2 \eta \boldsymbol{\mathsf{E}} + (\eta_b - \eta) \,  \mathrm{Tr}(\boldsymbol{\mathsf{E}}) \, \boldsymbol{\mathsf{I}}
+ \zeta \Delta \mu \; f(c) \; \boldsymbol{\mathsf{I}},
\end{align}
where the strain rate $\boldsymbol{\mathsf{E}} = \big[\nabla \boldsymbol{v} + (\nabla \boldsymbol{v})^{\mathsf{T}} \big]/2$ with $(\cdots)^{\mathsf{T}}$ denoting the transpose operation, $\eta$ and $\eta_b$ are respectively the shear and bulk viscosities of the cortex, $\mathrm{Tr}(\cdots)$ represents the trace operation, $\boldsymbol{\mathsf{I}}$ is the identity tensor, $\zeta \Delta \mu $ represents the strength of activity, and $f(c)$ is the active stress regulation function. Without loss of generality, we choose
\begin{align}
f(c) &= \frac{c}{c+c_s},
\end{align}
where $c_s$ is a saturation concentration. Combining \eqref{eqn_force_balance} and \eqref{eqn_total_stress}, the equation for the flow $\boldsymbol{v}$ is
\begin{align}
\label{eqn_velocity}
\eta \; \nabla^2 \boldsymbol{v} + \eta_b \; \nabla(\nabla \cdot \boldsymbol{v})  - \gamma \boldsymbol{v} = -\zeta \Delta \mu \;\nabla f(c).
\end{align}
We study \eqref{eqn_concentration} and \eqref{eqn_velocity} on an arbitrary two-dimensional domain $\Omega$. On the boundary $\partial \Omega$ of the domain, we impose a no-flux boundary condition for $c$ and a vanishing velocity condition:
\begin{align}
(\boldsymbol{\hat n} \cdot \nabla c) \big|_{\partial \Omega} = 0, 
\qquad
\boldsymbol{v} \big|_{\partial \Omega} = 0,
\label{eq:bc}
\end{align}
where $\boldsymbol{\hat n}$ is the outward unit normal in the domain. Since $f$ is a function of $c$, this implies that $(\boldsymbol{\hat n}  \cdot \nabla f) \big|_{\partial \Omega} = 0$. On the boundary $\partial\Omega$, the unit normal $\hat{\boldsymbol{n}}$ and the unit tangent $\hat{\boldsymbol{\tau}}$ are such that $\hat{\boldsymbol{n}} \cdot \hat{\boldsymbol{\tau}} = 0$ and we assume $\hat{\boldsymbol{n}} \times \hat{\boldsymbol{\tau}} = \boldsymbol{\hat z}$. We also define, respectively, the normal $\partial_{\boldsymbol{n}} \equiv \hat{\boldsymbol{n}} \cdot \nabla$ and the tangential derivatives $\partial_{\boldsymbol{\tau}} \equiv \hat{\boldsymbol{\tau}} \cdot \nabla$ on the boundary. We scale length, time and concentration by $\ell$, $\ell^2/\mathcal{D}$ and $c_0$ respectively where $\ell = \sqrt{\eta/\gamma}$. As such, the non-dimensional equations for $c$ and $\boldsymbol{v}$ are
\begin{align}
\partial_t c = -\nabla \cdot (\boldsymbol{v} c) + \nabla^2 c - R \; (c-1),
\label{eq:nondim_eqn_c}
\\
\nabla^2 \boldsymbol{v} + \frac{\eta_b}{\eta}  \nabla (\nabla \cdot \boldsymbol{v}) - \boldsymbol{v} = - P \; \nabla \left( \frac{c}{c+c_\ast}\right),
\label{eq:nondim_eqn_v}
\end{align}
where the Damk\"ohler number $R=\kappa \ell^2/\mathcal{D}$ compares the turnover timescale $\kappa^{-1}$ with the timescale for diffusion on a length-scale $\ell$, and the P\`eclet number $P=\zeta \Delta \mu/\gamma \mathcal{D}$ is the ratio of activity to diffusion. The other non-dimensional parameters are $c_\ast = c_s/c_0$, the ratio $\nu^2 = 1 + \eta_b/\eta$, and those associated with the size and shape of the domain $\Omega$. We note that the boundary condition $\boldsymbol{v}\vert_{\partial \Omega} = 0$ for the flow implies that \eqref{eq:nondim_eqn_v} uniquely determines $\boldsymbol{v}$ as a function of $c$. This can be seen by multiplying \eqref{eq:nondim_eqn_v} by $\boldsymbol{v}$ and integrating over the domain $\Omega$. 

\section{Effective model for the patterns}
\label{sec:framework}

In this section, we first develop a formalism to solve \eqref{eq:nondim_eqn_v} on any closed two-dimensional domain $\Omega$ of arbitrary shape. Using this, we setup a linear stability analysis calculation of the system \eqref{eq:nondim_eqn_c} and \eqref{eq:nondim_eqn_v} around the homogeneous state ($c=1, \boldsymbol{v}=0$). Next, we will develop a Galerkin truncated system of equations for the expansion coefficients of the concentration field. These equations will provide an effective low-dimensional model that, as we will demonstrate in \ref{sec:results}, predicts the phase-diagram of the concentration patterns even in the nonlinear regime.

Let $\psi_j$ be the orthonormal Neumann eigenfunctions of the Laplace operator on $\Omega$, i.e., $\nabla^2 \psi_j = -\lambda_j \psi_j$ with $\partial_{\boldsymbol{n}}\psi_j \vert_{\partial \Omega} = 0$, $\lambda_j \geq 0$, and $\int_{\Omega} \psi_j^\dagger \, \psi_k = \delta_{jk}$ where $\dagger$ represents complex conjugation. The $\psi_j$ form an orthonormal basis for expanding any function. Note that $j$ is, in general, a multi-index. Furthermore, the eigenfunction corresponding to the zero eigenvalue is a constant on $\Omega$ and $\int_\Omega \psi_j=0$ for $\lambda_j > 0$. In particular, we write
\begin{align}
c(\boldsymbol{x},t) = 1 + \sum_{k} c_k(t)  \, \psi_k(\boldsymbol{x}),
\label{eq:c_expansion}
\end{align}
where the summation is over the eigenfunctions $\psi_k$ with  eigenvalues $\lambda_k > 0$. We now multiply \eqref{eq:nondim_eqn_c}  by $\psi^\dagger_j(\boldsymbol{x})$ and integrate over $\Omega$ to obtain
\begin{align}
\label{eq:galerkin equation}
\frac{d c_j}{dt} = - (\lambda_j + R) \, c_j + \int_{\Omega} \nabla\psi_j^\dagger \cdot (\boldsymbol{v} c),
\end{align}
where we have performed an integration by parts and used the boundary condition $\boldsymbol{v}\vert_{\partial\Omega}=0$. If we can now express the flow $\boldsymbol{v}$ in terms of the concentration $c$, we will get a closed system of equations for the $c_j$. Note that this will lead to an infinite hierarchy of equations for the $c_j$. To obtain an effective reduced dynamical system, we shall truncate the expansion \eqref{eq:c_expansion} by retaining only a few dominant eigenfunctions. A phase-space analysis of this dynamical system will then yield a phase-diagram of the possible patterns.

For a given active stress function $f$ (equivalently, given $c$), an explicit solution of the flow equation \eqref{eq:nondim_eqn_v} is straightforward only in the case of simple rectangular domains. For boundaries with non-trivial geometry, it is a non-trivial task to obtain a solution for $\boldsymbol{v}$ satisfying \eqref{eq:nondim_eqn_v} with the boundary condition $\boldsymbol{v} \big|_{\partial \Omega} = 0$. In the next subsection, we outline a procedure that, in principle, allows us to compute the vector field $\boldsymbol{v}$ given the Green's function for the Dirichlet boundary-value problem for a scalar field satisfying a modified Helmholtz equation. Subsequently, we use this procedure to compute the flow field and then develop the Galerkin truncation for a disk shaped domain.

\subsection{Solving for the active flows}
Defining $\boldsymbol{C} \equiv \nabla \times \boldsymbol{v} = C \; \boldsymbol{\hat z}$ and $D \equiv \nabla \cdot \boldsymbol{v}$, the flow equation \eqref{eq:nondim_eqn_v} leads to
\begin{align}
\nabla^2  C - C = 0, 
\quad
\nu^2 \nabla^2 D - D = -P \nabla^2 f. 
\label{eqn_curl_div}
\end{align}
Using \eqref{eq:nondim_eqn_v}, we get $\boldsymbol{v}$ as
\begin{align}
\label{eqn_vel_D_C_f}
\boldsymbol{v} = \nu^2 \; \nabla D + \nabla_{\perp} C + P \; \nabla f,
\end{align}
where $\nabla_{\perp} \equiv \hat{\boldsymbol{z}} \times \nabla$. Using \eqref{eqn_vel_D_C_f}, the flow boundary conditions $\hat{\boldsymbol{n}} \cdot \boldsymbol{v}\vert_{\partial \Omega} = 0 = \hat{\boldsymbol{\tau}} \cdot \boldsymbol{v}\vert_{\partial \Omega}$ translate to the following coupled equations
\begin{align}
\nu^2 \partial_{\boldsymbol{n}} D -  \partial_{\boldsymbol{\tau}} C = 0,
\quad
\nu^2 \partial_{\boldsymbol{\tau}} D + \partial_{\boldsymbol{n}} C = -P \partial_{\boldsymbol{\tau}} f.
\label{eq:bc_D_C}
\end{align}
It is convenient to work with homogeneous equations. As such, we define
\begin{align}
\label{eq:newD_defn}
D = \mathsf{D} - P \sum_j \frac{\lambda_j \, f_j}{\nu^2 \lambda_j+1} \psi_j,
\quad\mathrm{with}\quad
f_j = \int_\Omega \psi_j^\dagger \, f.
\end{align}
Then $C$ and $\mathsf{D}$ satisfy the following homogeneous equations on $\Omega$
\begin{align}
\nabla^2  C - C = 0,
\quad
\nu^2\nabla^2 \mathsf{D} - \mathsf{D} = 0.
\label{eqn_C_D}
\end{align}
with the boundary conditions
\begin{align}
\nu^2 \partial_{\boldsymbol{n}} \mathsf{D} -  \partial_{\boldsymbol{\tau}} C = 0,
\quad
\nu^2 \partial_{\boldsymbol{\tau}} \mathsf{D} + \partial_{\boldsymbol{n}} C = F,
\label{eq:bc_C_D}
\end{align}
where
\begin{align}
F = -P \sum_j \frac{f_j \, \partial_{\boldsymbol{\tau}}\psi_j}{\nu^2\lambda_j+1}.
\end{align}
Note that in the above equation $\partial_{\boldsymbol{\tau}}$ is the tangential derivative evaluated on the boundary $\partial \Omega$. These boundary conditions \eqref{eq:bc_C_D} are neither in the Dirichlet form nor in the Neumann form. Nevertheless we show in Appendix \ref{appendix_uniqueness} that the boundary-value problem defined by \eqref{eqn_C_D}-\eqref{eq:bc_C_D} has a unique solution. 

To obtain a useful representation of the solution to \eqref{eqn_C_D}-\eqref{eq:bc_C_D}, we relate the solution to the Dirichlet boundary-value problem associated with \eqref{eqn_C_D}. To this end, suppose $\Omega$ is a subset of the plane such that $\partial\Omega$ is parameterized by twice-continuously differentiable function. Then equations \eqref{eqn_C_D} possess a unique solution when their respective Dirichlet conditions, i.e., $C|_{\partial\Omega}$ and $\mathsf{D}|_{\partial\Omega}$, are prescribed on $\partial\Omega$ \cite{han2011basic}. In particular, there exists an operator $\mathbb{L}$ (the `Green's function' for the Dirichlet boundary-value problem) such that $C = \mathbb{L} (C|_{\partial\Omega})$. Similarly $\mathsf{D}=\mathbb{L}_{\nu}(\mathsf{D}|_{\partial\Omega}$) with $\mathbb{L}_{\nu=1} = \mathbb{L}$. Moreover, there is a unique Neumann condition corresponding to the solution of the Dirichlet boundary-value problem. Let $\mathbb{G}$ denote the operator which maps the Dirichlet condition $C|_{\partial\Omega}$ to the associated Neumann condition $\partial_{\boldsymbol{n}} C$ when $C$ satisfies $\nabla^2 C - C = 0$. In other words, $\mathbb{G}(C|_{\partial\Omega}) = \partial_{\boldsymbol{n}} C$. Likewise $\mathbb{G}_{\nu}(\mathsf{D}|_{\partial\Omega})=\partial_{\boldsymbol{n}} \mathsf{D}$ with $\mathbb{G}_{\nu=1} = \mathbb{G}$. The operators $\mathbb{G}$ and $\mathbb{G}_{\nu}$ are known as Dirichlet-Neumann operators. These operators are self-adjoint and positive, hence invertible (see Appendix \ref{appendix_invertibility}). Further, we show in Appendix \ref{appendix_general_solution} that for a large class of domains $\Omega$, the solution to \eqref{eqn_C_D}-\eqref{eq:bc_C_D} can be expressed in the form
\begin{align}
\label{eq:finalCD}
C = \mathbb{B}_{\nu} F \;, 
\qquad
\mathsf{D} = \mathbb{A}_{\nu} F,
\end{align}
where $\mathbb{B}_{\nu} = \mathbb{L}\, \mathbb{K}_\nu^{-1} \, \mathbb{G}^{-1}$, $\mathbb{A}_\nu = \mathbb{L}_{\nu} \, \mathbb{G}_{\nu}^{-1} \, \partial_{\boldsymbol{\tau}} \mathbb{K}_\nu^{-1} \, \mathbb{G}^{-1}$ and $\mathbb{K}_\nu=\mathbb{G}^{-1} \, \partial_{\boldsymbol{\tau}} \, \mathbb{G}_{\nu}^{-1} \, \partial_{\boldsymbol{\tau}} + \mathbb{I}$ with $\mathbb{I}$ the identity operator. Note that $\mathbb{L}$, $\mathbb{G}$, $\mathbb{L}_{\nu}$, and $\mathbb{G}_{\nu}$ can be computed explicitly only for a few domains $\Omega \subset \mathbb{R}^2$. Despite the rarity of explicit solutions to \eqref{eqn_C_D}-\eqref{eq:bc_C_D} due to the nontrivial dependence on the shape of the domain $\Omega$, the cases where explicit solutions are available still provide valuable insight. The reason being that the flow $\boldsymbol{v}$ on a domain $\Omega$ is qualitatively similar to the flow on any domain $\Omega_{\epsilon}$ that is diffeomorphic to $\Omega$ (see Appendix \ref{appendix_general_solution}).

To summarize, using \eqref{eq:newD_defn} and \eqref{eq:finalCD} we get the formal expressions for the curl and divergence of the flow
\begin{align}
\label{eq:C_as_f}
C &= -P \sum_j \frac{\mathbb{B}_\nu \partial_{\boldsymbol{\tau}} \psi_j}{1+\nu^{2}\lambda_j} \, f_j,
\\
\label{eq:D_as_f}
D &= -P \sum_j \frac{\mathbb{A}_\nu \partial_{\boldsymbol{\tau}} \psi_j+ \lambda_j\psi_j}{1+\nu^{2}\lambda_j} \, f_j.
\end{align}
and the flow solution
\begin{align}
\boldsymbol{v} = -P \sum_j f_j \, \boldsymbol{u}_j(\boldsymbol{x}),
\label{eq:flow_operators}
\end{align}
with
\begin{align}
\boldsymbol{u}_j(\boldsymbol{x}) = \frac{\nu^{2} \nabla (\mathbb{A}_\nu \partial_{\boldsymbol{\tau}} \psi_j)
+  \nabla_{\perp} (\mathbb{B}_\nu \partial_{\boldsymbol{\tau}} \psi_j)
- \nabla \psi_j}{1+\nu^{2}\lambda_j}.
\end{align}
In the above expression, it should be noted that though $\partial_{\boldsymbol{\tau}} \psi_j$ are evaluated at the boundary $\partial\Omega$ of the domain $\Omega$, the result of the action of the operators $\mathbb{A}_{\nu}$ and $\mathbb{B}_{\nu}$  on $\partial_{\boldsymbol{\tau}} \psi_j$ is a function defined on $\Omega$. The gradients $\nabla$ and $\nabla_\perp$ then act on this function leading to the vector field $\boldsymbol{u}_j(\boldsymbol{x})$ defined on $\Omega$. The vector fields $\boldsymbol{u}_j(\boldsymbol{x})$ act as a basis for the expansion of the $\boldsymbol{v}$ with expansion coefficients $f_j$.

\subsection{Linear stabilty analysis and Galerkin truncation}
\label{sec:lsa:galerkin}

We  now use the formal solution for the flow \eqref{eq:flow_operators} in \eqref{eq:galerkin equation} to systematically analyze the resulting patterns in the concentration field $c(\boldsymbol{x},t)$.

To analyze the stability of the homogeneous state ($c=1, \boldsymbol{v}=0$), we consider $c(\boldsymbol{x},t) = 1 + \delta c(\boldsymbol{x},t) \equiv 1 + \sum_j \delta c_j(t) \, \psi_j(\boldsymbol{x})$ with $\delta c_j \ll 1$. It is easy to see from \eqref{eq:galerkin equation} that only the divergence $D$ of the flow contributes to the evolution of the $\delta c_j$ at the linear order. As such, the corresponding approximation for the expansion coefficient $f_j$ of the active stress regulation function $f(c)$ is given by
\begin{align}
f_j \approx \frac{c_\ast}{(1+c_\ast)^2} \delta c_j.
\end{align}
Using this approximation in \eqref{eq:D_as_f}, the linearized equations of motion for the $\delta c_j$ are 
\begin{align}
\label{eq:lsa_general_domain}
\frac{d \delta c_j}{dt} &= - (\lambda_j + R) \, \delta c_j + 
\frac{P \, c_\ast}{(1+c_\ast)^2} 
\bigg[
\frac{\lambda_j}{1+\nu^{2}\lambda_j} \delta c_j
\nonumber \\ & \qquad
+
\sum_k \frac{1}{1+\nu^{2}\lambda_k} \left(\int_{\Omega} \psi_j^\dagger  \mathbb{A}_\nu \partial_{\boldsymbol{\tau}} \psi_k \right) \, \delta c_k
\bigg].
\end{align}
The integral involving $\mathbb{A}_\nu$ and the eigenfunctions $\psi_j$ depends on the geometry of $\Omega$ via the Green's function $\mathbb{L}$ and the Dirichlet-Neumann operator $\mathbb{G}$. The evolution equations for the $\delta c_j$ are coupled to each other. In other words, the linear stability matrix is non-diagonal in the $\psi_j$ basis. This turns out to be a generic feature for most two-dimensional domains. Note that \eqref{eq:lsa_general_domain} holds for all $\delta c_j$ under the assumption that they are small.

Next, we turn to develop a systematic Galerkin truncation of the modes in the expansion \eqref{eq:c_expansion}. We first consider $c_{\ast} \gg 1$ and approximate $f(c) \approx c (c_\ast - c)/c_\ast^2$. We then use the expansion \eqref{eq:c_expansion} for $c$, multiply the approximate expression for $f$ by an eigenfunction $\psi_j^\dagger$ and integrate over $\Omega$ to get
\begin{align}
\label{eq:fj_galerkin}
f_j \approx \frac{(c_\ast - 2)}{c_\ast^2} \, c_j -  \frac{1}{c_\ast^2} \sum_l \sum_m c_l c_m \, \int_\Omega \psi_j^\dagger \, \psi_l \, \psi_m.
\end{align}
We now insert the above approximation for $f_j$ in \eqref{eq:flow_operators} and use the resulting expression for $\boldsymbol{v}$ in the evolution equations \eqref{eq:galerkin equation} along with the expansion \eqref{eq:c_expansion}. This leads to
\begin{align}
\label{eq:galerkin_general_domain}
\frac{d c_j}{dt} &= - (\lambda_j + R) \, c_j 
- \frac{P}{c_\ast^2} 
\bigg[ 
(c_\ast - 2) \, \sum_k  \mathcal{E}_{jk} \, c_k
\nonumber \\ & \qquad
+ \sum_{k,l} \big[ (c_\ast - 2) \mathcal{F}_{jkl} - \sum_n \mathcal{E}_{jn} \mathcal{I}_{nkl} \big] \, c_k c_l
\nonumber \\ & \qquad
-  \sum_{k,l,m} \big[\sum_n  \mathcal{F}_{jkn}  \mathcal{I}_{nlm}\big] \, c_k c_l c_m
\bigg],
\end{align}
where the expansion coefficients are
\begin{subequations}
\begin{align}
\mathcal{E}_{jk} &= \int_{\Omega} \nabla\psi_j^\dagger \cdot \boldsymbol{u}_k,
\\
\mathcal{F}_{jkl} &= \int_{\Omega} \psi_k (\nabla\psi_j^\dagger \cdot \boldsymbol{u}_l),
\\
\mathcal{I}_{lmn} &= \int_\Omega \psi_l^\dagger \, \psi_m \, \psi_n,
\end{align}
\label{eq:galerkin_integrals}
\end{subequations}
These integrals depend solely on the geometry of the domain $\Omega$ (the only material parameter is $\nu$ entering via the operators $\mathbb{A}_\nu$ and $\mathbb{B}_\nu$). 

In principle, equations \eqref{eq:galerkin_general_domain} provide an almost exact and closed set of equations for the time-evolution of the amplitudes $c_j$ in \eqref{eq:c_expansion} -- the only approximation being \eqref{eq:fj_galerkin}. However, equations \eqref{eq:galerkin_general_domain} are an infinite hierarchy of coupled nonlinear equations for the $c_j$. As such, to get an effective low-dimensional dynamical system, we will necessarily have to truncate the expansion by retaining only a few $c_j$. Further, even after this truncation, obtaining an analytically tractable system requires that we are able to evaluate the integrals in \eqref{eq:galerkin_integrals} on the domain $\Omega$. This requires that we know the Neumann eigenfunctions $\psi_j$ of the Laplacian operator, the Green's functions $\mathbb{L}$, $\mathbb{L}_{\nu}$ for the Dirichlet boundary value problem, as well as the Dirichlet-Neumann operators $\mathbb{G}$, $\mathbb{G}_{\nu}$. As remarked earlier, analytical expressions for these quantities are possible only in very few cases. However, they can always be computed numerically for any geometry.

\section{Results}
\label{sec:results}

In this section, we present results on the steady-state patterns of concentration results from \eqref{eq:nondim_eqn_c} and \eqref{eq:nondim_eqn_v} on a general two-dimensional domain $\Omega$. We parametrize the domain $\Omega$ with a generalized set of orthogonal coordinates $\{\xi_1, \xi_2\}$ such that the boundary $\partial \Omega$ is defined as the level set $\xi_1 = a$, where $a$ is a dimensionless number. Following \cite{JENKINS_DYSTHE_1997}, we will use $\eta_b/\eta=3$ (and hence $\nu = 2$) unless stated otherwise. First, we shall illustrate the  formalism developed in the previous section in the case when $\Omega$ is a disk domain of radius $a$. The phases resulting from both linear-stability analysis and the Galerkin truncation will be compared with numerical solutions of \eqref{eq:nondim_eqn_c} and \eqref{eq:nondim_eqn_v}. Next, we will consider domains that are smooth deformations of the disk and illustrate how boundary curvature controls the localization of concentration patterns.

\subsection{Patterns on a disk}

\begin{figure}[t]
\centering    \includegraphics[width=\linewidth]{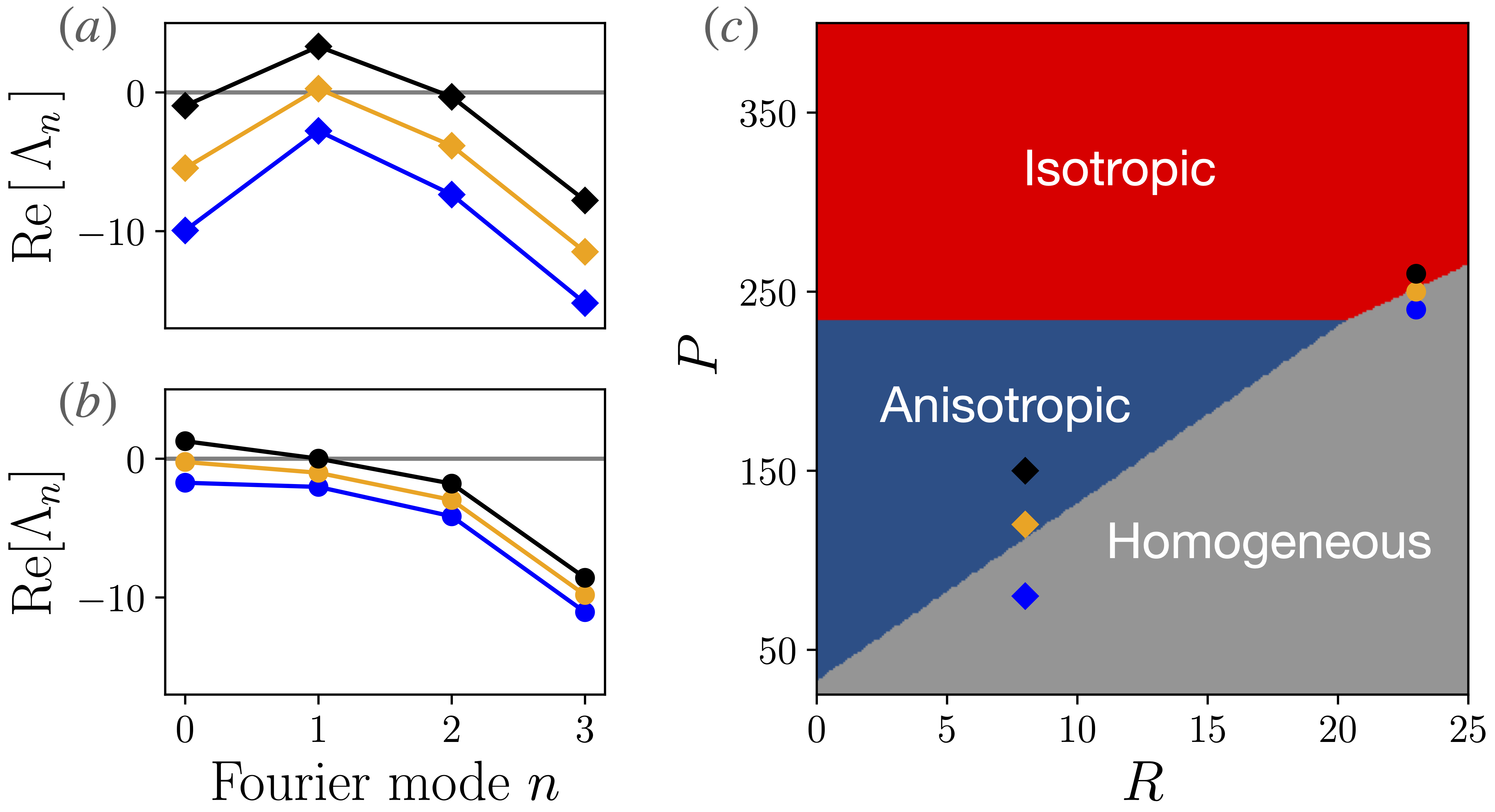} 
\caption{Linear stability analysis on a disk-shaped domain. Upon increasing $P$, the homogeneous state loses its stability with either (a) anisotropic modes having the largest growth rate at small $R$, or (b) isotropic modes having the largest growth rates at large $R$. The parameter values corresponding to the dispersion curves in (a) and (b) are marked in (c). The other parameters are fixed at $a=2, c_\ast=4$. From the phase-diagram in the $P-R$ plane shown in (c), we conclude that linear stability predicts the pattern will always isotropize at large $P$.}
\label{fig:lsa_circle}
\end{figure}

On any domain $\Omega$, the Dirichlet values $C|_{\partial\Omega}$ and $\mathsf{D}|_{\partial\Omega}$ can be expanded in a suitable orthonormal basis of functions $\{u_n\}$ defined on $\partial\Omega$. The choice of the orthonormal basis $\{u_n\}$ depends on the shape of the domain $\Omega$. On the unit disc, a natural choice is $u_n = (2\pi)^{-1} e^{in\theta}$ for $\theta\in[0,2\pi]$ and $n\in\mathbb{Z}$. Note the tangential-derivative operator equals $\partial_\theta$ in this basis and hence is a diagonal operator. The $u_n$ form an orthonormal basis for the space of square integrable functions on the circle denoted by $L^2(\partial\Omega)$. For such a choice, we note that 
\begin{align}
\langle u_m,\partial_{\boldsymbol{\tau }}u_n\rangle_{L^2(\partial\Omega)} = \int_0^{2\pi} \frac{d\theta}{2\pi} e^{-im\theta}\partial_\theta e^{in\theta} = in\delta_{mn}.
\end{align}
On a disk of radius $a$, the normalized Neumann eigenfunctions of the Laplacian are
\begin{align}
\psi_{n,k}=\frac{1}{N_{n,k}}\, J_n \left(\alpha_{n,k} \frac{r}{a}\right)e^{in\theta}
\end{align}
where $J_n$ is the $n^{\rm th}$ order Bessel function of the first kind, $n\in\mathbb{Z}$, the $\alpha_{n,k}$ satisfy  $J_n'(\alpha_{n,k})=0$ where $'$ denotes differentiation,  $k=1,2,3,\ldots$, and the constant $N_{n,k}$ is set so that $\psi_{n,k}$ has unit $L^2-$norm. The eigenvalue corresponding to $\psi_{n,k}$ is $\lambda_{n,k} = \alpha_{n,k}^2$.
Further
\begin{subequations}
\begin{align}
\mathbb{A}_\nu e^{in\theta} &= \frac{in}{\nu} \,\frac{I_n(a) I_n(r/\nu)}{d_n} \,e^{in\theta},
\label{eq:A_operator_disk}
\\
\mathbb{B}_\nu e^{in\theta} &= \frac{I_n'(a/\nu) I_n(r)}{d_n} \, e^{in\theta},
\label{eq:B_operator_disk}
\end{align}
\end{subequations}
where $d_n=I_n'(a)I_n'(a/\nu)-n^2I_n(a)I_n(a/\nu)$ and $I_n$ is the $n^{\rm th}$ order modified Bessel function of the first kind.

With this setup, we can now proceed towards the linear stability analysis of the patterns. We can explicitly evaluate the integral in \eqref{eq:lsa_general_domain} using \eqref{eq:A_operator_disk} above. As remarked earlier, the linear stability matrix in the case of the unit-disc also turns out to be non-diagonal in the $\psi_{n,k}$ basis. The equations for the perturbation amplitudes $\delta c_{n,k}$ remain coupled with respect to the index $k$. To proceed, we truncate to the first ten eigenfunctions in the $k$-index, and then for each $n$  compute the largest eigenvalue $\Lambda_n$ of the linear stability matrix. Fig~\ref{fig:lsa_circle}(a-b) show the variation of $\Lambda_n$ with $n$ at several parameter values. The growth rates seen in Fig~\ref{fig:lsa_circle}(a) show that the homogeneous state becomes unstable with increasing $P$ and transitions into an anisotropic pattern ($n\neq0)$. On the other hand, figure~\ref{fig:lsa_circle}(b) shows that homogeneous state loses its stability and transitions into an isotropic pattern ($n=0)$. This allows us to construct the linear stability phase-diagram in the $P-R$ plane for patterns in the disk-shaped domain. We observe that the homogeneous state can lose its stability, upon increasing  $P$, to either an anisotropic pattern or an isotropic pattern depending on the value of the turnover rate $R$. This linear stability analysis predicts that isotropic patterns can arise at large $P$ even when $R=0$.

\begin{figure*}[ht]
\includegraphics[width=\textwidth]{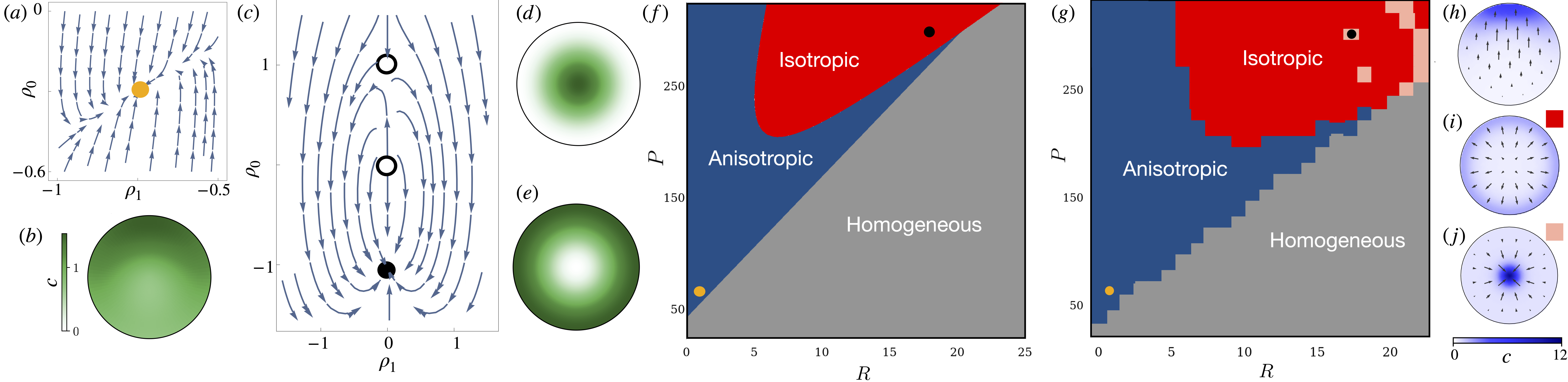}
\caption{Comparison of the results from the truncated model (a-f) in \eqref{eq:galerkin_equation_disk} with numerical solution of the full model (g-j) in \eqref{eq:nondim_eqn_c}-\eqref{eq:nondim_eqn_v} for a disk-shaped domain with size $a=2$ and $c_\ast=4$. (a) Phase portrait of the system \eqref{eq:galerkin_equation_disk} where the stable fixed point leads to the anisotropic concentration pattern shown in (b). The phase is arbitrarily set to $\varphi_1=\pi/2$. (c) Isotropic fixed points ($\rho_0 \neq 0$ and $\rho_1=0$) lead to concentration patterns shown in (d) or (e). The Galerkin truncated model \eqref{eq:galerkin_equation_disk} predicts that the fixed point corresponding to the pattern in (d) is unstable while that corresponding to (e) is stable. We also see that the homogeneous fixed point ($\rho_0=0=\rho_1$) is unstable. (f) Phase diagram showing stability regions of the homogeneous, isotropic and anisotropic fixed points from the Galerkin model. (g) Phase diagram resulting from the numerical integration of \eqref{eq:nondim_eqn_c}, \eqref{eq:nondim_eqn_v}. (h-j) Concentration patterns corresponding to parameter values marked in (g).
}
\label{fig:galerkin}
\end{figure*}

Next, we proceed to develop an effective low-dimensional Galerkin model for the patterns. We truncate the concentration field as $c(r,\theta,t) = 1 + \rho_0 \, \psi_{0,1} + \rho_1 \, e^{i\varphi_1} \, \psi_{1,1} - \rho_1 \, e^{-i\varphi_1} \, \psi_{-1,1}$ where the amplitudes $\rho_i \in \mathbb{R}$ and the phase $\varphi_1 \in \mathbb{R}$. Using this truncation in \eqref{eq:galerkin_general_domain} and evaluating the integrals \eqref{eq:galerkin_integrals}, we get
\begin{subequations}
\label{eq:galerkin_equation_disk}
\begin{align}
\dot{\rho}_0 &= - (R + \alpha_{0,1}^2)\rho_0 
- \frac{P}{c_\ast^2} \Big[
(c_\ast-2) a_{01} \rho_0 
\nonumber \\
& \qquad
- (c_\ast-2) (a_{02} \rho_0^2 - a_{04} \rho_1^2)
+ a_{03} \rho_0^2 - a_{05} \rho_1^2 
\nonumber \\
& \qquad
+ a_{06} \rho_0^3 - a_{07}\rho_0 \rho_1^2
\Big],
\\
\dot{\rho}_1 &= - (R + \alpha_{1,1}^2) \rho_1 + \frac{P}{c_\ast^2} \Big[(c_\ast-2) a_{11} \rho_1 
\nonumber
\\
& \qquad 
- [(c_\ast-2) a_{12} + a_{13}]\rho_0 \rho_1  
\nonumber
\\
& \qquad
+ a_{14} \rho_1 \rho_0^2 - a_{15} \rho_1^3 \Big],
\end{align}
\end{subequations}
and $\dot{\varphi}_1 = 0$. Here the coefficients $a_{pq}>0$ are related to the integrals in \eqref{eq:galerkin_integrals} and their values are listed in Appendix~\ref{appendix_disk_values}. We immediately see that the phase $\varphi$ is decoupled from $\rho_i$ and can be chosen arbitrarily with the choice fixing the direction corresponding to $\theta=0$.  

We summarize the results of the patterns arising from the Galerkin truncated model \eqref{eq:galerkin_equation_disk} in Fig~\ref{fig:galerkin}. Analyzing \eqref{eq:galerkin_equation_disk}, we find three kinds of fixed-points: (i)$\rho_0 = \rho_1 =0$,  (ii) $\rho_0 \neq 0$, $\rho_1 = 0$, (iii) $\rho_0 \neq 0$ and $\rho_1  \neq 0$.  These respectively correspond to the homogeneous state, an isotropic pattern, and an anisotropic pattern. Within isotropic patterns ($\rho_1=0$), the concentration could be localized at the boundary ($\rho_0 < 0$) or localized at the center of the domain ($\rho_0 > 0$). Fig~\ref{fig:galerkin}(a) shows the phase-portrait of the system around the fixed-point corresponding to the anisotropic pattern. We also plot, in Fig~\ref{fig:galerkin}(b), the resulting concentration pattern arising from the fixed-point values for $\rho_i$. In Fig~\ref{fig:galerkin}(c), we display the phase-portrait of the system around the fixed-points corresponding to the two isotropic patterns. We clearly see that the fixed-point corresponding to $\rho_0>0$ is unstable while that corresponding to $\rho_0<0$ is stable. Fig~\ref{fig:galerkin}(d-e) show the respective concentration patterns at these fixed-points. We  proceed to analyze the existence and the linear stability of the fixed points of \eqref{eq:galerkin_equation_disk} as a function of the parameters. In Fig~\ref{fig:galerkin}(f), we construct the resulting phase-diagram of the patterns in the $P-R$ plane. The parameter values corresponding to the anisotropic and isotropic patterns shown in Fig~\ref{fig:galerkin}(b,d,e) are indicated in the phase-diagram. We notice from this phase-diagram that the Galerkin truncated model predicts that isotropic patterns do not occur when $R=0$, i.e., when there is no turnover of the concentration $c$. This should be contrasted with the prediction from the linear stability analysis of the homogeneous state shown in Fig~\ref{fig:lsa_circle}.

To validate the predictions of the Galerkin truncated model, we numerically solve the full model \eqref{eq:nondim_eqn_c}-\eqref{eq:nondim_eqn_v} with the finite-element method using the FEniCS library~\cite{dolfin_GNwells, ufl_GNwells, mainfenics_GNwells}. For each parameter set, we start from the homogeneous state $c=1$ and $\boldsymbol{v}=0$ with small random perturbations added to $c$ and time-evolve the equations until a steady-state is reached. The results of this explicit numerical calculation are summarized in the phase-diagram Fig~\ref{fig:galerkin}(g). The concentration and flow patterns corresponding to the parameter values indicated in the phase-diagram are shown in Fig~\ref{fig:galerkin}(h-j). We observe the following points: (i) we find a very good agreement between the patterns predicted from the Galerkin model \eqref{eq:galerkin_equation_disk} (and Fig~\ref{fig:galerkin}(f)) and those seen from the explicit numerical solution of the partial differential equations for $c$ and $\boldsymbol{v}$, (ii) the transition between the anisotropic and isotropic patterns (occuring in the highly nonlinear regime) predicted by the Galerkin model is captured quite well in the full numerics, (iii) consistent with the prediction of the Galerkin model, we do not get isotropic patterns when $R=0$, (iv) in contrast to the prediction of the Galerkin model, we see a stable coexistence of two kinds of isotropic patterns (shown in Fig~\ref{fig:galerkin}(i-j)) at high values of $P$ and $R$, and finally (v) the phase diagram from the Galerkin model indicates that anisotropic patterns will remerge for large values of $P$ when $R$ is small; this is not seen in the full numerics. It is remarkable that a low-dimensional Galerkin model retaining only two modes can account for a large number of features seen in Fig~\ref{fig:galerkin}(g)).

\subsection{Deformations of the disk}

\begin{figure}[ht]
\centering   \includegraphics[ width=\linewidth]{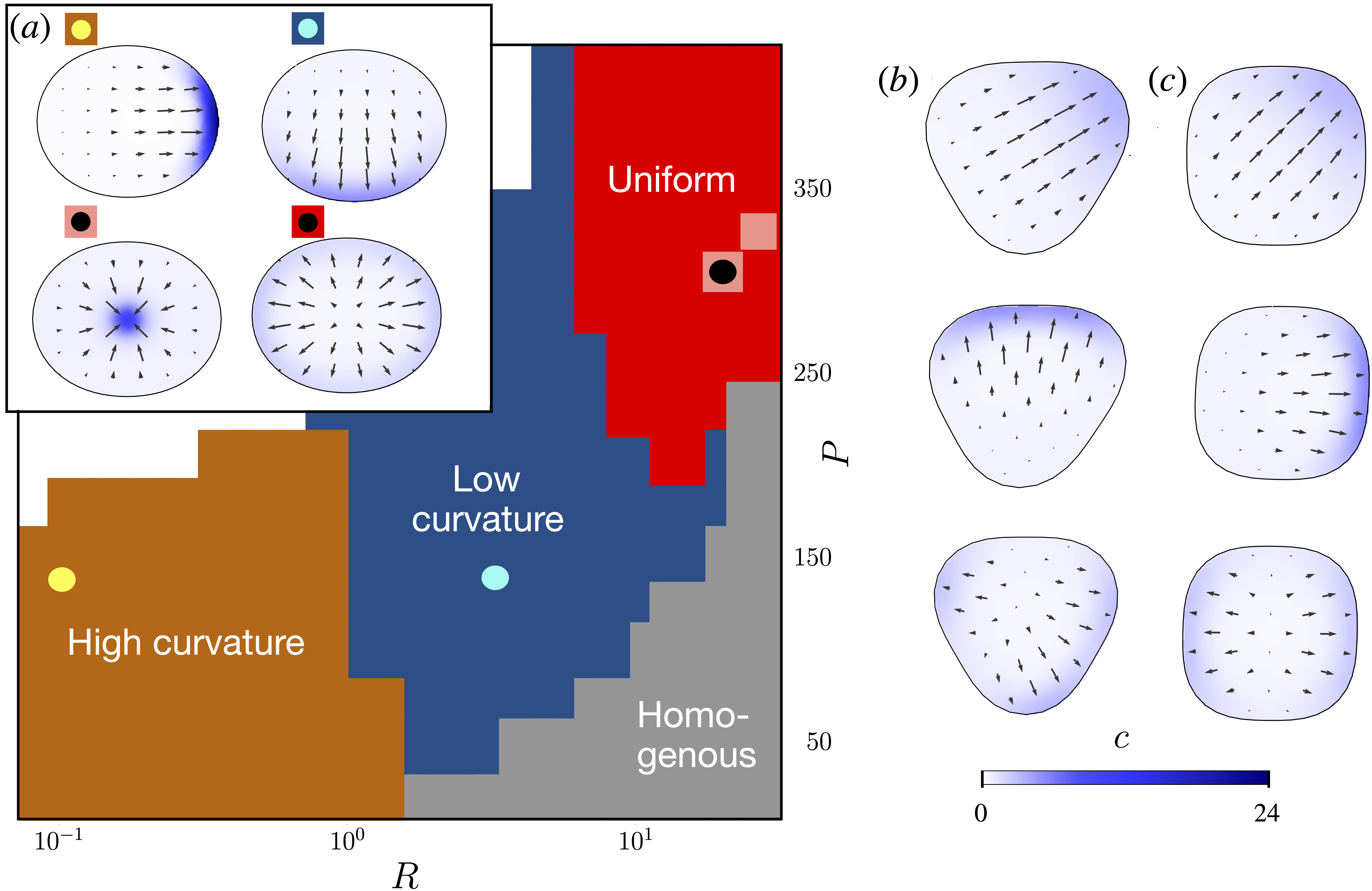}
\caption{(a) Anisotropic patterns on an $l=2$ deformation of the circle either select regions of high or low boundary curvature. At high values of $P$ and $R$, the patterns localize either at the center of the domain or along the boundary. The phase-diagram in the $P-R$ plane is also shown. Patterns on (b) $l=3$  and (c) $l=4$ deformations can localize to vertices, edges, or both. In (b) and (c), the parameter values are ($P=50, R=1$), ($P=150, R=5$), and ($P=200, R=9$) for each row. Here $a=2$ and $c_\ast=4$.}
\label{fig:numerics_noncircular}
\end{figure}

The formalism that we developed in Section \ref{sec:framework} applies to any (piecewise) smooth two-dimensional domain. However, analytical calculations are possible only on those domains for which the eigenfunctions of the Laplace operator, the Green's function for the Dirichlet problem and associated operators can be explicitly obtained. In the previous section, we showed that the concentration patterns, and the transitions between them, on a disk shaped domain can be accounted by a low-dimensional Galerkin model. To explore how the shape of the domain affects the resulting steady state concentration patterns, we consider smooth perturbations of the disk parametrized by 
\begin{align}
r(\theta)=a\left(1+\sum_{l=2}^\infty {r}_l \cos(l\theta)\right),
\end{align}
where $r_l$ is the amplitude of deformation corresponding to the $l$-th mode.

We first consider deformations with $r_l = 0, \, \forall \, l>2$. For small $r_2$, this results in a convex shape with a varying boundary curvature. To explore concentration patterns on such a domain, we numerically integrate \eqref{eq:nondim_eqn_c}-\eqref{eq:nondim_eqn_v} to find steady state solutions. The results of this calculation are summarized in the phase diagram in Fig~\ref{fig:numerics_noncircular}(a). We observe that the concentration patterns either localize nearly uniformly on the boundary or select regions of high/low boundary-curvature. For the case of the disk-shaped domain, the location of the peak of the anisotropic pattern was decided based on the random initial conditions. In contrast, on a domain deformed with an $l=2$ mode, we find that the concentration patterns localize to regions of high boundary-curvature at low values of $P$ and $R$. At intermediate values of $P$ and $R$, the patterns transition to localize at regions of low boundary-curvature. Further, at large values of $P$ and $R$, we find that there is a stable coexistence of concentration patterns that either spread nearly uniformly throughout the boundary, or are localized in the middle of the domain.

\begin{figure}[hb]
\centering
\includegraphics[width=\linewidth]{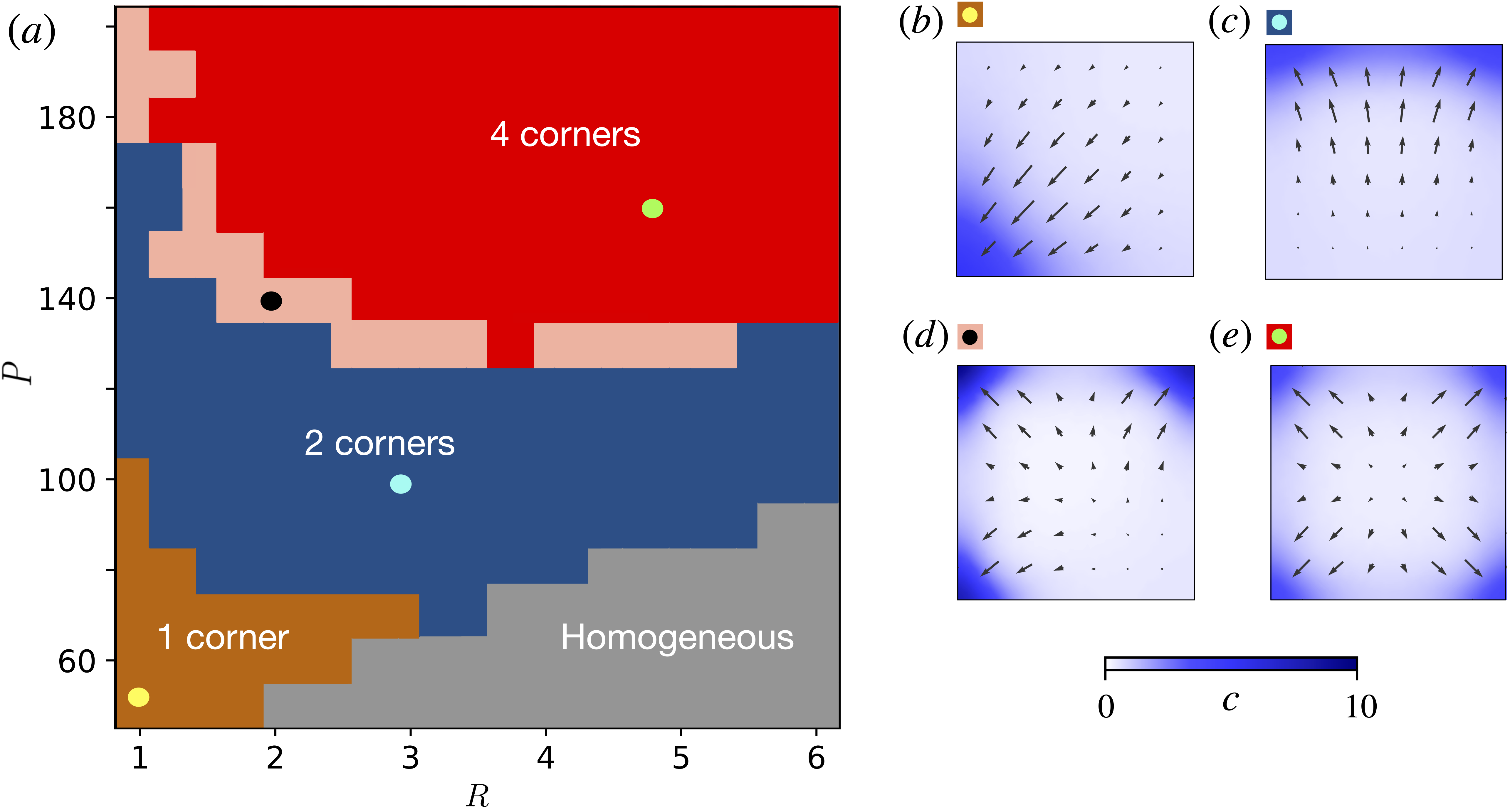}
\caption{Patterns on a square localise at one, two, three or four corners (shown respectively in (b)-(e)). The parameter values for (b)-(e) are marked in the phase-diagram shown in (a). Here $a=2, c_\ast=4$.}
\label{fig:square_phase_diagram}
\end{figure}

A shape closely related to the $l=2$ deformation of the disk is an ellipse. However when $\Omega$ is an ellipse, in spite of the fact that separation of variables in elliptic coordinates provides explicit solutions to both the Laplace and Helmholtz equations, the Dirichlet-Neumann  and  tangential-derivative operators cannot be simultaneously diagonalized. As a consequence, one does not obtain explicit solutions to the boundary-value problem \eqref{eqn_C_D}-\eqref{eq:bc_C_D} for this choice of $\Omega$. Nevertheless, we have numerically solved \eqref{eqn_C_D}-\eqref{eq:bc_C_D} on a ellipse-shaped domain and find that the resulting patterns is very similar to those shown in Fig.~\ref{fig:numerics_noncircular}(a).

For shapes with higher mode deformations of the disk, we expect a plethora of rich concentration patterns. For instance, on a convex domain with only an $l=3$ deformation, the steady state patterns can either localize to a vertex, to an edge, or be localized at a vertex and the opposite edge, as shown in Fig~\ref{fig:numerics_noncircular}(b). On a smooth square-shaped domain, concentration patterns can localize to single vertices, single edges, or to opposite edges (Fig~\ref{fig:numerics_noncircular}(c)).

The derivation of the truncated Galerkin equations may be carried out when $\Omega$ is a rectangular domain aligned with the coordinate axes in $\mathbb{R}^2$. Note, since the rectangular domain is \emph{not} diffeomorphic to the circle (due to the presence of sharp corners in the rectangular domain), one does not expect the associated velocity fields to be qualitatively similar. Indeed, one expects the solutions to differ precisely at the corner points. Nevertheless there are some striking similarities between the concentration patterns obtained for the smooth square-shaped domain (Fig~\ref{fig:numerics_noncircular}(c)) and the sharp-cornered square domain (Fig~\ref{fig:square_phase_diagram}). In either case, solutions localize to the boundary regions. Further investigation is needed regarding the precise role the domain-geometry (and specifically the role of corners) plays in determining fixed-points of the truncated Galerkin equation and their stability properties.

\section{Discussion}
\label{sec:discussion}

In this study, we studied pattern formation in a hydrodynamic model for the actomyosin cortex of cells. In addition to the physical parameters of the system, we have developed a systematic framework to investigate the influence of the boundary geometry in controlling the emergent patterns. For a disk-like geometry, our framework leads to explicit equations for the amplitudes of a Galerkin mode-truncated system. We demonstrated that the steady-state solutions of this reduced dynamical system captures the features of the patterns seen in the full partial differential equation system. The phase-diagram of patterns in the $P-R$ plane predicted by the Galerkin model shows a remarkable agreement with that obtained from an explicit numerical calculation of the full model. On the disk-shaped domain, our results predict that anisotropic patterns that break azimuthal symmetry can bifurcate into isotropic patterns with changing parameter values. On domains that are harmonic deformations of the disk, our numerical analysis shows that the emergent patterns can select regions of high or low boundary curvature depending on the parameter values.

We observe the following points. First, our model equations \eqref{eq:nondim_eqn_c} and \eqref{eq:nondim_eqn_v} contain a very minimal mechanochemical coupling between the concentration field $c$ and the flows $\boldsymbol{v}$ resulting from the active stresses. What we have explored is the role of boundary geometry in controlling the resulting patterns. Our analytical framework for developing a Galerkin truncated low-dimensional model can be extended in a straightforward manner to include multiple interacting chemical species \cite{BoisPRL2011, KumarPRL2014}. Second, the sensitivity of the patterns to the boundary geometry arises from the independent manner in which advective and diffusive transports affect the dynamics of $c$. As we have shown, the solution of the flow equation \eqref{eq:nondim_eqn_v} in an arbitrary domain with Dirichlet boundary conditions is given in terms of quantities involving Dirichlet-Neumann operators acting on velocity gradients. In other words, stresses developed at the boundaries can have a strong influence on the flow profiles, and hence on the concentration patterns. Third, the subtle differences between advective and diffusive transport is apparent even at the linear stability level. As mentioned in Sec~\ref{sec:lsa:galerkin}, the linear stability matrix is not diagonal in the basis of the eigenfunctions $\psi_j$ of the Laplace operator. The reason is that the Laplace and advection operators \emph{do not} commute with each other in our case. This should be contrasted with the same problem in periodic and no-flux domains in one-dimension \cite{BoisPRL2011}, or on the sphere \cite{mietkeMinimalModelCellular2019}. Fourth, on domains with varying boundary curvature (such as the $l=2$ deformation of the disk shown in Fig~\ref{fig:numerics_noncircular}), the combined action of active advection and diffusion leads to patterns localizing at regions of the boundary with high or low curvature. In fact, we notice at a given high value of $P$, the boundary localization of patterns transitions from regions of high boundary curvature to regions of low boundary curvature as $R$ is increased. Notice that $R$ is itself the ratio of the length-scales associated with flow ($\ell = \sqrt{\eta/\gamma}$) and turnover ($\sqrt{D/\kappa}$). Varying boundary curvature introduces yet another length scale and the resulting competition leads to transitions in the localization of the pattern on the boundary.



Our study shows that a simple model for actomyosin patterns is sensitive to the geometry of the confining boundary. The predictions of our model can be tested in experiments where cells are plated on micropatterned substrates wherein the geometry of the confining region can be easily controlled. Further directions could explore including an orientational order parameter such as the nematic alignment of actin fibers and also explore domains that can deform in response to the boundary stresses.

\begin{acknowledgments}

We acknowledge support of the Department of Atomic Energy, Government of India, under project number RTI4001. SP thanks ICTS for support during the S.N. Bhatt Memorial Excellence Fellowship Program 2019.

\end{acknowledgments}

\bibliography{references}

\appendix

\section{Uniqueness of the boundary value problem for $C$ and $\mathsf{D}$}
\label{appendix_uniqueness}
As discussed in the main text, the curl $C$ and the (modified) divergence $\mathsf{D}$ of the flow field $\boldsymbol{v}$ satisfy the following homogeneous equations on $\Omega$
\begin{align}
\nabla^2  C - C = 0,
\quad
\nu^{2} \;\nabla^2 \mathsf{D} - \mathsf{D} = 0.
\label{apx_eqn_C_D}
\end{align}
with the boundary conditions
\begin{align}
4 \partial_{\boldsymbol{n}} \mathsf{D} -  \partial_{\boldsymbol{\tau}} C = 0,
\quad
4 \partial_{\boldsymbol{\tau}} \mathsf{D} + \partial_{\boldsymbol{n}} C = F.
\label{apx_bc_C_D}
\end{align}
To see that these equations have a unique solution, we first define a norm for a vector-field $\boldsymbol{v}$ defined on $\Omega$ in the following way
\begin{align*}
\| \boldsymbol{v}\|^2 = \int_{\Omega}\left[\mathrm{tr}(\nabla\boldsymbol{v}\nabla\boldsymbol{v}^T)+ (\nu^2-1)(\nabla\cdot\boldsymbol{v})^2+\boldsymbol{v}\cdot\boldsymbol{v}\right]\:
\end{align*}
Then the solution to \eqref{apx_eqn_C_D}-\eqref{apx_bc_C_D} when $F=0$, satisfies the following
\begin{align*}
\| \nu^{2}\nabla \mathsf{D} + \boldsymbol{\hat z}\times \nabla C \|^2 = 0\:.
\end{align*}
Then from the definition of the norm, we conclude $\mathsf{D} = 0\:, C = \mbox{constant}$ in $\Omega$ for such a solution. But the only constant solution to \eqref{apx_eqn_C_D} is zero. Thus the only solution to \eqref{apx_eqn_C_D}-\eqref{apx_bc_C_D} with zero boundary forcing, is the zero solution. Hence the boundary-value problem has a unique solution.

\section{Invertibility of the Dirichlet-Neumann operator}
\label{appendix_invertibility}

Using Green's identity, one can show that the Dirichlet-Neumman operator is a self-adjoint operator acting on square-integrable functions defined on $\partial\Omega$ \cite{lannes2013water}. Moreover, the Dirichlet-Neumann operator for modified Helmholtz is an invertible operator which can be seen from the following version of Green's identity obtained from the equation for $C$ in \eqref{apx_eqn_C_D}
\begin{align}
\int_{\partial\Omega}C|_{\partial \Omega}\, \mathbb{G}(C|_{\partial\Omega}) = \int_{\Omega}(|\nabla C|^2 + \mu^2 C^2)\:.
\end{align}
Indeed from the well-posedness of the Dirichlet boundary-value problem,  the right-hand side above vanishes if and only if $C|_{\partial\Omega}=0$. Hence $\mathbb{G}$ (and $\mathbb{G}_\nu$) is positive-definite and hence invertible.

\section{Solving for  $C$ and $\mathsf{D}$}
\label{appendix_general_solution}

The boundary-conditions \eqref{apx_bc_C_D} may be re-expressed in the following matrix-operator equation
\begin{align}
\begin{pmatrix}
4\mathbb{G}_{\nu}    & -\partial_{\boldsymbol{\tau}} \\
4\partial_{\boldsymbol{\tau}} & \mathbb{G}
\end{pmatrix}  
\begin{pmatrix}
\mathsf{D}_{\partial\Omega}\\ C_{\partial\Omega}
\end{pmatrix}
= 
\begin{pmatrix}
0 \\ F
\end{pmatrix}
\label{eq:bc_CD_DNO}
\end{align}
As a consequence of the invertibility of $\mathbb{G}$ proved above, we can rewrite \eqref{eq:bc_CD_DNO} in the following equivalent form
\begin{align}
\label{eq:bc_CD_scalar}
( \mathbb{G}^{-1}\partial_{\boldsymbol{\tau}}\mathbb{G}_{\nu}^{-1}\partial_{\boldsymbol{\tau}} +I )C|_{\partial\Omega} &= \mathbb{G}^{-1}F\\
\mathsf{D}|_{\partial\Omega} &= \frac{1}{4} \mathbb{G}_{\nu}^{-1}\partial_{\boldsymbol{\tau}}C|_{\partial \Omega} 
\end{align}
Recall the boundary-value problem \eqref{apx_eqn_C_D}-\eqref{apx_bc_C_D} has a unique solution. Thus the matrix-operator on the left-hand side of \eqref{eq:bc_CD_DNO} and the scalar-operator on the left-hand side of \eqref{eq:bc_CD_scalar} are both invertible. By expanding  $C|_{\partial\Omega}$ and $\mathsf{D}|_{\partial\Omega}$ in a suitable orthonormal basis of functions $\{u_j\}$ defined on $\partial\Omega$, \eqref{eq:bc_CD_scalar} becomes a linear system of equations for the respective expansion coefficients. Owing to the uniqueness of solutions, the expansion coefficients in the ${u_j}$-basis are uniquely determined from this system of equations. As a consequence, the Dirichlet conditions $C|_{\partial\Omega}$ and $\mathsf{D}|_{\partial\Omega}$ are uniquely determined and moreover the solutions to \eqref{apx_eqn_C_D} given by $\mathbb{L}(C|_{\partial\Omega})\:,\mathbb{L}_\nu(\mathsf{D}|_{\partial\Omega})$ are also uniquely determined. Consequently, we obtain the velocity field $\boldsymbol{v}$.

The above calculations can be carried out explicitly when $\Omega$ is the unit disc. For smooth deformations of the unit-disc given by
\begin{align}
r(\theta) = 1 + \epsilon h(\theta)\quad \theta\in[0,2\pi]\:,
\label{appx_disc_perturb}
\end{align}
one can, in principle, exploit the analyticity of the Dirichlet-Neumann operator $\mathbb{G}$ to obtain solution to \eqref{eq:bc_CD_scalar} perturbatively. For the above to be a diffeomorphism of the unit circle, it suffices to demand $|\epsilon h'|\leq c < 1$ for some real number $c$. It is well-known that the Dirichlet-Neumman operator depends smoothly (indeed analytically) on the shape of the domain. In particular this implies $\mathbb{G}, \, \mathbb{G}_{\nu}$ (and their inverses) depend analytically on $\epsilon$, i.e., these operators can be written as a power-series in $\epsilon$ with coefficients as self-adjoint operators \cite{lannes2013water}. Though it is tedious to obtain explicit forms of the $\epsilon-$expansions for $\mathbb{G}, \, \mathbb{G}_{\nu}$ (see \cite{bruno2001boundary,nicholls2001new}) we can however conclude that the solution to  \eqref{apx_eqn_C_D}-\eqref{apx_bc_C_D} on the domain given by \eqref{appx_disc_perturb} is qualitatively close to the solution of the same boundary-value problem but posed on the unit disc.

\section{Integrals in the Galerkin truncated equations for a disk}
\label{appendix_disk_values}

We tabulate below the values of the coefficients in the Galerkin truncated model for patterns on the disk.\\

\begin{minipage}{\textwidth}
\begin{tabular}{| c  c  c  c  c  c  c |}
\hline
$a_{01}$ & $a_{02}$ & $a_{03}$ & $a_{04}$ & $a_{05}$ & $a_{06}$ & $a_{07}$ \\ 
\hline
$0.936$ & $0.165$ & $0.33$ & $0.865$ & $0.455$ & $0.121$ & $0.519$ \\ 
\hline
\end{tabular}

\vspace{5pt}

\begin{tabular}{| c  c  c  c  c |}
\hline
$a_{11}$ & $a_{12}$ & $a_{13}$ & $a_{14}$ & $a_{15}$ \\
\hline
$0.633$ & $0.117$ & $0.199$ & $0.049$ & $0.037$ \\
\hline
\end{tabular}
\end{minipage}

\end{document}